\documentclass{aa}

\usepackage{graphics}
\usepackage{longtable}

\begin{document}

\title{Fine structure of the chromospheric activity \\ in Solar-type stars --  The H$\alpha$ line
\thanks{Based on observations collected at Observat\'orio do Pico dos Dias,
operated by the Laborat\'orio Nacional de Astrof\'{\i}sica, CNPq, Brazil.}
}

\author{W. Lyra \&  G. F. Porto de Mello}

\offprints{W. Lyra}

\institute{
UFRJ/Observat\'orio do Valongo, Ladeira do Pedro Ant\^onio, 43\\
20080-090 Rio de Janeiro, RJ, Brazil\\
emails: wlyra@ctio.noao.edu,gustavo@ov.ufrj.br}

\date{Received 12 February 2004 / Accepted 07 October 2004}

\authorrunning{Lyra \& Porto de Mello}
\titlerunning{H$\alpha$ Chromospheric Emission}

\abstract{ A calibration of H$\alpha$ as both a chromospheric diagnostic and an age indicator is presented, complementing the works previously done on this subject (\cite{herbig}, \cite{luca1}). The chromospheric diagnostic was built with a statistically significant sample, covering nine years of observations, and including 175 solar neighborhood stars. Regarding the age indicator, the presence of stars for which very accurate ages are determined, such as those belonging to clusters and kinematic groups, lends confidence to our analysis. We also investigate the possibility that stars of the same age might have gone through different tracks of chromospheric decay, identifying - within the same age range - effects of metallicity and mass. These parameters, however, as well as age, seem to be significant only for dwarf stars, losing their meaning when we analyze stars in the subgiant branch. This result suggests that, in these evolved stars, the emission mechanism cannot be magnetohydrodynamical in nature, in agreement with recent models (Fawzy et al. 2002c, and references therein). The Sun is found to be a typical star in its H$\alpha$ chromospheric flux, for its age, mass and metallicity. As a byproduct of this work, we developed an automatic method to determine temperatures from the wings of H$\alpha$, which means the suppression of the error inherent to the visual procedure used in the literature.


    }

\maketitle

\section{Introduction}

Losing angular momentum through magnetized stellar winds, cool main sequence dwarfs have their rotation continuously braked, reducing the efficiency of their dynamos and, consequently, their degree of chromospheric activity.  Because of this, the chromospheric filling observed in high opacity spectral lines can be translated into a potential indicator of age, a quantity which is still one of the most uncertain parameters in stellar astrophysics.

Although this scenario has a remarkable simplicity, it is a subject which has remained largely unexplored in a quantitative way. Among the works which have been published, most tend to focus on the H and K lines of Ca\,{\small {II}} (e.g. \cite{skumanich}, \cite{linsky}), in which the chromospheric emission is more obvious, but also more affected by transient phenomena and phase modulation. The H$\alpha$ line, although widely used to measure chromospheric activity in solar physics, has received less attention in relation to this particular problem, and the lack of a good calibration of H$\alpha$ is a drawback for several reasons. First, the line is less sensitive to transient phenomena like flares, coronal mass ejections and localized magnetic explosions; extremely energetic phenomena that flood the X-ray and ultraviolet spectra with energy, but barely affect the visible. Second, it has the property of characterizing the mean chromospheric flux in a better way than Ca\,{\small {II}} H and K because, showing less chromospheric filling, phase modulations within an activity cycle are greatly reduced: the errors in computing the flux -- due, for instance, to normalization and determination of effective temperatures --, largely overcome the intrinsic modulation. Third, the modern solid state detectors have higher quantum efficiency in the red, behaving inversely to the old photographic plates. Also, the studied stars --- solar type ones ---  have their maximum flux in the visible region, favoring better accuracy in narrow band photometry centered on H$\alpha$.

Nevertheless, even if H$\alpha$ did not present any advantage, the need for another diagnostic is crucial. As said before, the majority of works which attacked the problem until the 80s analyzed only the calcium lines, considering them representative of the radiative losses in the chromosphere. The core of H$\alpha$, however, is formed in different regions (Schoolman 1972), thus responding differently to changes in the physical conditions of the chromosphere. In this sense, calibrating H$\alpha$ will help not just to better determine the energy budget, but also to better discern the structure of upper stellar atmospheres. With this goal, Herbig (1985, hereafter H85) and Pasquini \& Pallavicini (1991, hereafter P$^2$91) presented the only works measuring the net chromospheric fluxes in H$\alpha$, finding consistent results using two different photometric wavebands to calibrate the spectra into absolute flux. Of these two attempts, one (H85) went so far as to develop an age indicator; however, the small sample of 43 stars used, in which all but two are field stars, did not allow for a sufficiently accurate analysis.

\begin{table*}
\begin{center}
\caption[]{Parameters of the three analyses of chromospheric activity on H$\alpha$ to date}
\label{param}
\begin{tabular}{c c c c}\hline

            & Herbig    & Pasquini \& Pallavicini   & This work         \\
            & 1985      & 1991              & 2004          \\ \hline
Resolution in the center of H$\alpha$ ({\AA})   & 0.74          & 0.11              & 0.30          \\
N$^{\d{o}}$ of stars        & 43        & 87            & 175           \\
Range of Temperature Classes    & F8-G3     & F8-K5         & F5-K0         \\
Luminosity Classes  & V         & IV \& V           & IV \& V       \\
Photometric Waveband    & Johnson's V & Willstrop's $\lambda\lambda$\,6650-6600\,{\AA} & Willstrop's $\lambda\lambda$\,6650-6600\,{\AA}\\
Age Indicator       & Yes       & No                & Yes           \\ \hline

\end{tabular}
\end{center}
\end{table*}

In this work, we intend to fill this gap, calibrating H$\alpha$ as an absolute diagnostic of chromospheric activity, using a statistically significant sample of 175 solar neighborhood stars, and building an age indicator based on this diagnostic, calibrated with stars belonging to open clusters and stellar kinematic groups (SKG). We also use a large sample of field stars with accurate ages derived from theoretical isochrones. This large sample is unique in the literature, and allows for a better statistical analysis. For comparison, the parameters of the three analyses on the H$\alpha$ chromospheric flux are shown in Table~\ref{param}. This paper is divided as follows: in section 2 we describe the observations and reduction procedures. In the 3$^{\rm rd}$ we build the chromospheric diagnostic with the purpose of, in section 4, calibrating the age indicator. In section 5 we investigate the influence of other parameters in the degree of chromospheric activity, outlining the results and proceeding to the conclusion in the 6$^{\rm th}$ section.

\section{Observations and reduction}
To provide a chromospheric flux vs. age calibration, field stars alone are not adequate, as ages determined either by isochrones or lithium abundances are not accurate enough; cluster stars are needed. Thus, we observed two open clusters -- Pleiades and Hyades -- and three Kinematic Groups - Ursae Majoris, $\zeta$ Reticuli{\footnote {The name by which this group, previously known as $\zeta$ Herculis, should be now known, according to \cite{peloso}}} and HR\,1614, besides a large sample of field stars. A kinematic group is a group of stars which presents the same galactic velocity components, evidence of a common origin. These stars thus form an HR diagram typical of stars of the same age and chemical composition, showing reduced scatter --- the same properties of a cluster, except for the fact that they do not present its spatial compactness. For these reasons, the SKG is believed to be the intermediate state of the dispersion process of a cluster into field stars (\cite{dave2}).

Stars belonging to the large on-going survey of Mt. Wilson (\cite{baliunas}) were also observed, not to constrain ages, but because their rotational periods and cycles of chromospheric activity have been monitored for 25 years, and some chromospherically quiet stars have been identified in this extensive database. These stars are important in determining the photospheric flux to be subtracted. All stars were selected to be observed in the southern hemisphere, with a magnitude limit of $V=8.5$, ensuring a high signal-to-noise ratio (250, on average) while keeping a moderately high resolution ($R=20\,000$).

Spectra were taken with the Coud\'e spectrograph coupled to the 1.60m telescope at Observat\'orio do Pico dos Dias (OPD, Bras\'opolis), operated by Laborat\'orio Nacional de Astrof\'{\i}sica (LNA/CNPq) in 13 runs (Sep/94, Feb/95, Jul/98, Sep/99, May/00, Oct/00, Jun/01, Sep/01, Oct/01, May/02, Aug/02, Oct/02, Dec/02). Starlight was passed through a 250 $\mu$m slit over a diffraction grating of 1800 l/mm in first direct order, and projected over the 1024 x 1024 pixels CCD camera, corresponding to a linear projection of 0.30{\AA} per pixel in the center of H$\alpha$. For the faintest stars in our sample, Pleiades' G dwarfs, integration times of 900s resulted in spectra of $S/N=30$. As at least ten of these spectra would be necessary for a combined spectrum of moderate S/N, the slit was broadened to 500 $\mu$m, worsening the resolution, but improving the S/N for the same exposure time. We compared the final fluxes taken with both slits, for a few bright stars, concluding that a good agreement was achieved. A total of 175 solar types stars were observed, a substantial improvement when compared to the analyses previously done by H85 and P$^2$91, with 45 and 83 stars, respectively (Table 1).

Reduction was carried out by the standard procedure using IRAF. After the usual bias and flat field correction, the background and scattered light were subtracted and the one dimensional spectra were extracted. The pixel/wavelength calibration was performed selecting isolated spectral lines from the Kurucz Solar Atlas (1984).  All spectra had their own wavelength scale, removing the need for Doppler shift correction. The rms achieved on H$\alpha$ centering was 3\,m{\AA} on average, being 6 m{\AA} in the noisiest spectra and less than 1\,m{\AA} in the best ones. Normalization of the continuum was performed fitting a low order polynomial in the regions not affected by H$\alpha$ profile. The unusable regions were visually determined in $\lambda\lambda$\,6515-6600\,{\AA}, a large fraction of the total coverage of our spectra, $\lambda\lambda$\,6485-6630\,{\AA}. All spectra were taken with the same instruments and reduced by the same person, which makes this analysis homogeneous.

\section{Building the chromospheric diagnostic}

\subsection{Computing the net chromospheric fluxes}

\begin{figure}
\begin{center}
\resizebox{10cm}{7cm}{\includegraphics{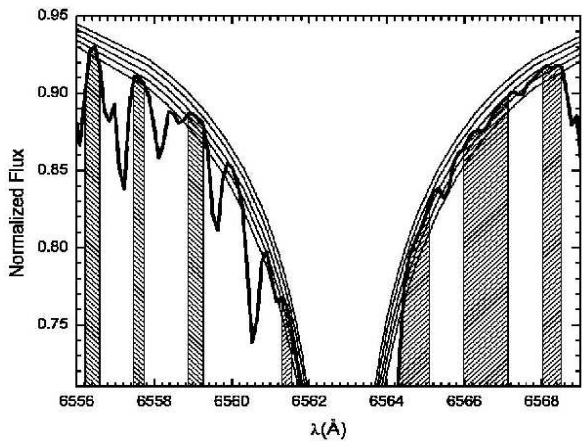}}
\end{center}
\caption[]{Sketch of the automated method to determine effective temperatures from the wings of H$\alpha$. The five adjacent curves are five theoretical profiles representing wings formed in temperatures spaced by 50\,K intervals. The shadowed intervals are, as determined by inspecting Kurucz et al.'s solar atlas (\cite{kurucz}), free from any metallic lines. These intervals are then used to determine the temperatures, pixel by pixel. The effective temperature is computed as 5782\,$\pm$\,19\,K. Occasional telluric features and eventual weak metal lines not present in Kurucz's solar atlas are removed by Kolmogorov-Smirnov statistics.}
\label{wings}
\end{figure}

\begin{figure}
\resizebox{10cm}{7cm}{\includegraphics{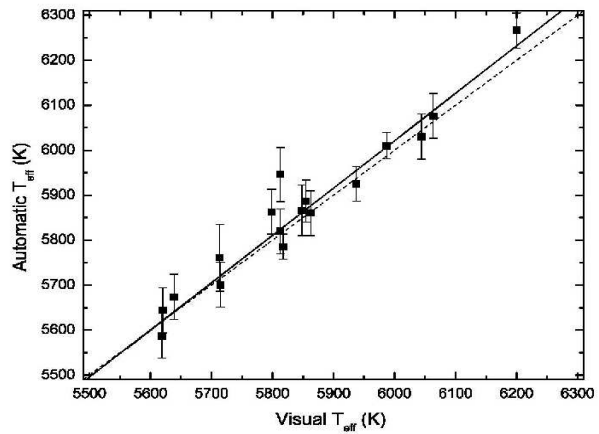}} \caption[]{Test of the automated process over some temperatures visually determined. The best fit (full line) is consistent with T$_{\rm aut}$ = T$_{\rm vis}$ (dashed line) within the range of one standard deviation.} 
\label{teffs}
\end{figure}

In order to compute the chromospheric losses, in erg\,cm$^{\rm -2}$\,s$^{\rm -1}$, on the stellar surface, one must: {\em a}. determine which width around the core of H$\alpha$ is to be used to measure the emission; {\em b}. calibrate the spectra in units of absolute flux on the surface of the star, in erg\,cm$^{\rm -2}$\,s$^{\rm -1}$\,{\AA}$^{\rm -1}$;  and {\em c}. remove the photospheric contribution.

A method to arrive at the chromospheric flux was developed by P$^2$91, based on the pioneering work of Linsky et al. (1979) in the Ca\,{\small{II}} lines. We use the same method in our analysis, which consists of measuring the area under 1.7\,{\AA} around H$\alpha$, transforming it in absolute fluxes at the surface of the star by means of the narrow photometric waveband, from 6550\,{\AA} to 6600\,{\AA}, calibrated by Willstrop (1965). Their method is briefly summarized in the following paragraphs.

From the familiar equation relating the flux measured on the Earth ($f$) and the absolute flux at the stellar surface ($F$)

\begin{equation}
F(\Delta\lambda) = f(\Delta\lambda)(D/R){^2},
\end{equation} where {\it D} stands for the stellar heliocentric distance and {\it R} for its radius, P$^2$91 used the Barnes-Evans relation (Barnes \& Evans 1976) and Willstrop's calibration to arrive at a relationship between absolute flux and the $(V-R)$ color index. They define this relationship as

\begin{equation}
\log F(50) = -1.4430 $(V-R)$ + 7.602,
\end{equation} where F(50) is the average flux at the stellar surface in the $\lambda\lambda$\,6650-6600\,{\AA} waveband, in erg\,cm$^{\rm -2}$\,s$^{\rm -1}$\,{\AA}$^{\rm -1}$. The error in the best fit of this calibration is estimated by the authors as 0.038 in the angular coefficient.

W$_{H\alpha}$ being the area under the central 1.7 {\AA} around H$\alpha$ core and W$_{50}$ the total area under Willstrop's waveband, the H$\alpha$ absolute flux at the surface of the star is calculated as

\begin{equation}
 F_{H\alpha} = {\frac{W_{H\alpha}}{W_{50}}}F(50)50.
\end{equation} where the factor 50 corresponds to the length of the waveband. It is needed because $F(50)$ is not the total flux, but this flux averaged in wavelength. Finally, P$^2$91 perform the photospheric correction by subtracting the lower boundary that arises from a $F_{H\alpha}$ vs. $(V-R)$ plot, assuming this boundary to represent the photospheric flux.

We followed strictly along the same lines, with only minor variations. A slight difference in our analysis was that we did not work with color indexes, but directly with effective temperatures, derived from the wings of H$\alpha$. In such a case, in order to use P$^2$91's eq.\,2, one has to calibrate a transformation between the two quantities, $(V-R)$ and {\em $T_{\rm{eff}}$}. Porto de Mello (1996) defines this transformation as

\begin{equation}
T_{\rm eff} = 8465 - 5055$(V-R)$,
\end{equation} valid for solar type stars, in the intervals from 5\,000 to 6\,500\,K in {\em $T_{\rm{eff}}$}  and from -1.0 to +0.30 in [Fe/H]. The quantities W$_{H\alpha}$, W$_{50}$ and F$_{H\alpha}$ are respectively shown in columns 10, 11 and 12 of Table~5 {\footnote {only available in electronic form at the CDS via anonymous ftp to cdsarc.u-strasbg.fr (130.79.128.5) or via http://cdsweb.u-strasbg.fr/cgi-bin/qcat?J/A+A/vol/page}}.

We remark that these temperatures were derived using a software we wrote, and not by the visual procedure used in the literature. This software compares the observed spectrum with theoretical ones, synthesized using the package AHYDRO, developed by Praderie (1967), assigning a temperature value for each pixel assumed free from metallic lines (Fig.~\ref{wings}), giving us a temperature distribution, a mean and a standard deviation. This distribution is Gaussian unless there is contamination by telluric features and eventual weak metal lines not present in the solar spectrum, both resulting in temperatures higher than expected.

To remove this contamination, the software tests if the residuals follow a Gaussian distribution, executing the Kolmogorov-Smirnov test on them. If a non-Gaussian distribution is found, the hottest pixel is removed and the test performed again, until it converges to a Gaussian. The derived temperatures are shown in the 5$^{\rm th}$ column of Table 5. The temperature achieved for the sun was 5782\,K, in excellent agreement with the accepted value of 5777\,K (Neckel 1986). Fig.~\ref{teffs} shows that the visual and automatic procedures agree within one standard deviation.

The methods agree because they both give reliable temperatures. The point that makes the automated procedure better is because it is, of course, completely systematic. The visual procedure is full of subjective decisions, as is any fit by eye. The agreement is always qualitative: fits by eye done by the same person on different occasions will give different temperatures. We made tests that consist of repeating the same eye-fitting several times, which convinced us that the derived temperatures oscillate within a range of 20\,K. This source of error, that we call {\it personal error}, is eliminated in an automated procedure.

We have also calculated the errors incurred on the temperature determination by other sources. So, errors in $\log g$, [Fe/H] and microturbulence of 0.25 dex, 0.15 dex and 0.2 km/s, which are representative for our sample, result in errors in the temperature determination of 25\,K, 10\,K and 10\,K. We have further estimated the uncertainty resulting from an error of 0.2\% in the continuum determination of H$\alpha$ spectra, a value obtained by us by normalizing different spectra of the same star obtained in different runs: this error is 25\,K. An {\it rms} composition of these contributions lead to a total estimated error of $\sim$50\,K in our temperatures.

\begin{figure}
\resizebox{10cm}{7cm}{\includegraphics{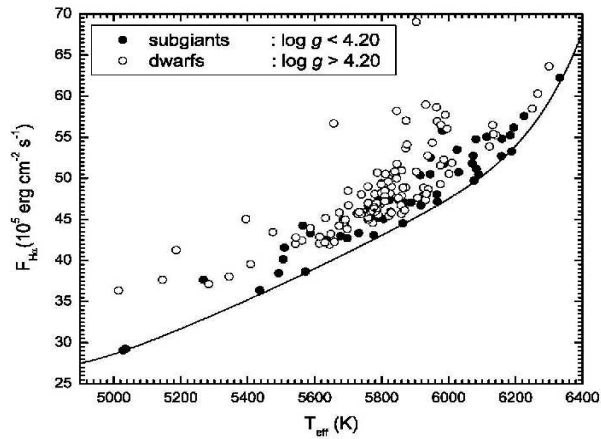}} \caption[]{A lower boundary of quiet stars arises from a $F_{H\alpha}$ vs. $T_{\rm{eff}}$ plot, defining the photospheric flux to be subtracted. The sample divided in dwarfs and subgiants, the two populations are seen to be clearly distinct, the subgiants being less active, on average.} 
\label{boundary}
\end{figure}

Another difference lies in the way to evaluate the width around H$\alpha$'s center to measure the chromospheric flux. As claimed by  P${^2}$91, the chromospheric profile seems to broaden as the star evolves. To verify this hypothesis, we separated the sample into dwarfs and subgiants, using the turnoff point as the boundary between the two kinds of stars. We calculated for several evolutionary tracks the surface gravity corresponding to the turnoff point, deciding that $\log g$ = 4.20 best defines it. Using this value as a boundary, we found that in the dwarfs the emission peak emcompasses 1.4\,{\AA} of the core of H$\alpha$; a value that rises to 1.6\,{\AA} in the subgiant branch. We stress, however, that these values are means, and a quite large scatter is present in both dwarfs and subgiants. As the greatest width found in our sample was  1.7\,{\AA}, we adopted this value to measure the chromospheric filling in H$\alpha$ ensuring that no chromospheric flux was left unregarded. Both H85 and P$^2$91 use the same value, by independent arguments.

We perform the subtraction of the photospheric contribution in the same way, tracing a natural spline in the lower boundary of quiet stars that arise from the flux\,-\,temperature plot. As shown in fig.~\ref{boundary}, this lower boundary is clearly defined by the subgiants. This result is expected, because the rotational velocity diminishes as the radius grows in the evolution from dwarf to subgiant, reducing the efficiency of the dynamo. As calculated by Fawzy et al. (2002c, and references therein), there is the possibility that the dynamo gets so weak that the tiny acoustic heating becomes more relevant.

Independently of the kind of star that defines the boundary, it defines the flux to be subtracted for each temperature. One must keep in mind, however, that this procedure is accompanied by a certain amount of arbitrariness, in the sense that it depends on the sample used. Moreover, the exact curve of photospheric flux might perhaps be impossible to determine. One can only know the {\em exact} photospheric flux if dealing with a star of absolutely no chromospheric flux. Such a star, however, does not exist, given that the mechanisms generating the dynamo are never completely turned off. Also, it is theorized that the basal acoustic heating in the low chromosphere is active even in hypothetical stars with no magnetohydrodynamical activity.

A way to solve this problem is to use another indicator in which the photospheric flux in the center is zero or nearly zero -- Ca\,{\small {II}} H and K or Mg\,{\small {II}} h and k -- and force the chromospheric flux of H$\alpha$ to go to zero with them. The drawback of this procedure lies in the fact that these lines modulate with the activity cycle, in such a way that simultaneous observations must be performed, largely multiplying the already tough observational demands. To make things worse, the formation mechanisms of the lines are different. A phenomenon that excites Ca\,{\small {II}} or Mg\,{\small {II}} transitions will not necessarily affect H$\alpha$, or vice-versa.

With all the possible effects combined, a relation $F'_{K} \,vs\, F'_{H\alpha}$ would eventually present considerable scatter, as shown in the analyses of both H85 and P$^2$91. In these works, a {\em negative} chromospheric flux was found for some stars after the photospheric correction by the Ca\,{\small {II}} K line, resulting from the inconvenient scatter in the linear regression. In this work, we did not perform simultaneous observations in Ca\,{\small {II}} K and H$\alpha$ for a large number of stars. In this case, to avoid worsening the homogeneity in determining the chromospheric component, as well as the occurrence of negative fluxes, we made the decision to accept the arbitrary nature of the method of subtracting the lower boundary of quiet stars.

\begin{figure}
\begin{center}
\resizebox{7cm}{!}{\includegraphics{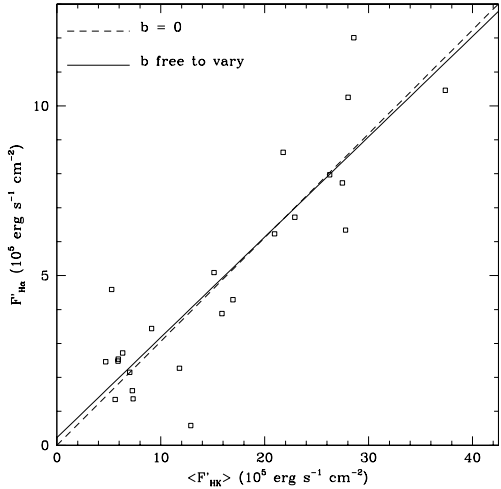}}
\end{center}
\caption[]{Least squares fit of the H$\alpha$ flux upon the flux in the calcium lines averaged in the cycle of activity. No significant difference is found between the fit done allowing the linear coefficient to vary (solid line) and constraining it to zero (dashed line). It indicates that both quantities indeed go to zero together and no correction by the calcium lines is needed.}
\label{calcium}
\end{figure}

However, to investigate whether the correction with the calcium line is actually needed, we can plot the quantity thus derived, $F'_{H\alpha}=F_{H\alpha}-F_{phot}$, against the flux in the calcium lines {\em averaged} in the cycle of activity. Baliunas et al. (1995), provide accurate HK measurements over 25 years for 111 stars, 33 of them being present in our sample.

We used the calibration derived by Noyes et al. (1984) to transform the index $<${\em S}$>$ into absolute fluxes. With these fluxes, we arrive at the relations shown in figure~\ref{calcium}. In this figure, we show least square fits done allowing the linear coefficient  {\em b} to vary (solid line) and constraining it at zero (dashed line). The difference is negligible, which means that both quantities go to zero together. The quantity $F''_{H\alpha}$, defined by P$^2$91, does not seem to be needed.

\begin{figure}
\begin{center}
\resizebox{7cm}{7cm}{\includegraphics{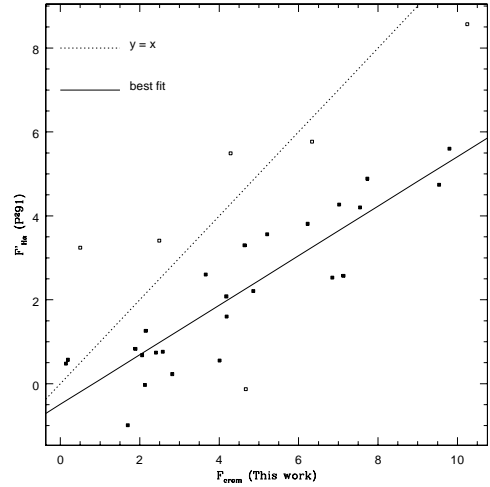}}
\caption[]{Comparison between our fluxes and those obtained by P$^2$91. The dashed line represents the curve y=x, the ideal agreement. The best fit (solid line) shows that the agreement is, in fact, far from ideal: our fluxes are systematically greater than those of P$^2$91 (see text). Open squares refer to points that lie above or below the 2$\sigma$ limit of the fit.} 
\label{comparison}
\end{center}
\end{figure}
 
We compared our fluxes, $F_{crom}$, with those obtained by P$^2$91, $F'_{H\alpha}$, using 31 stars in common between the two samples (fig ~\ref{comparison}). Embarassingly, not only is a large scatter present, but our fluxes are systematically greater than theirs. As we followed the same procedure, a likely explanation for this discrepancy lies in the different ways to express the quantity F(50). P$^2$91 calibrate this quantity using $(V-R)$, but for many stars this index was not available, and so they used transformations to get $(V-R)$ from $(B-V)$. Their temperature scale being photometric in nature, we have investigated the possibility of systematic differences between this scale and our H$\alpha$ one. For many stars there exist photometric determinations of temperature: we have checked these against the H$\alpha$ temperatures and no systematic difference emerged. At present, we cannot offer an explanation for the systematic difference between our fluxes and P$^2$91's, though we note that the scatter is considerable.

\subsection{Estimate of the errors}

\begin{figure}
\resizebox{10cm}{7cm}{\includegraphics{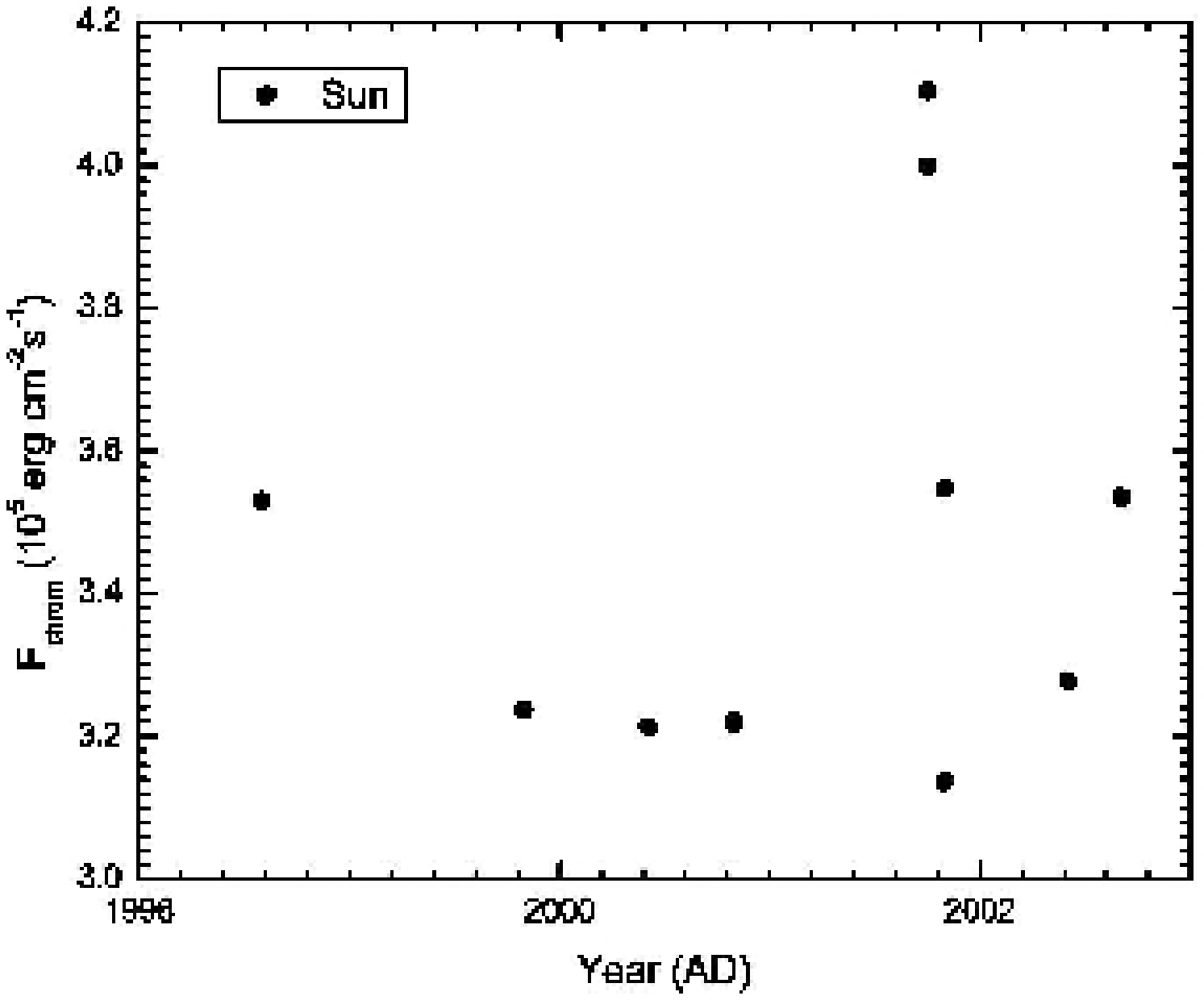}}
\resizebox{10cm}{7cm}{\includegraphics{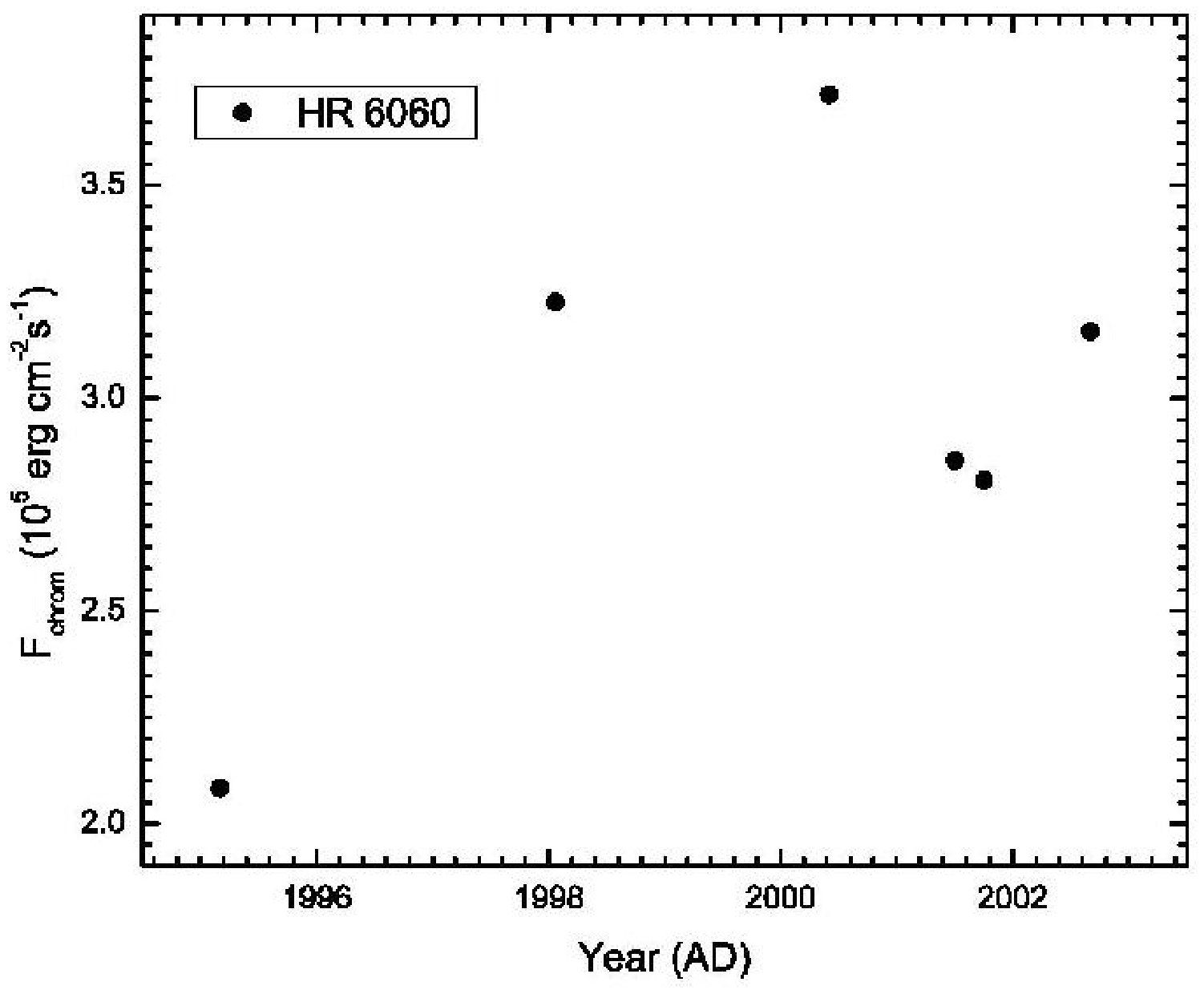}} \caption[]{Our measurements of the activity level of the Sun and HR\,6060, during the last solar cycle, having maximum in June 2000. The large scatter observed is evidence that H$\alpha$ better characterizes the mean level of activity. The scatter shown in the graphs seems not to be physical, but rather statistical.} \label{cycle}
\end{figure}

The error in the chromospheric fluxes were derived by repeated observations of the same star. Spectra of the Sun and HR\,6060 were taken each 6 months or nearly so, with the purpose of characterizing a solar activity cycle and, using this cycle as a template, to investigate the existence of a similar cycle in HR\,6060, believed to be the closest ever solar twin star
(\cite{gemea}; \cite{gemea2}).

The solar data (fig.~\ref{cycle}), however, do not allow us to recognize a definite cycle. The maxima and minima observed do not match the physical ones either. We are led to believe that the scatter in this graph is not physical, and is likely just statistical fluctuation.

We also investigated the occurrence of rotational modulation, a phenomenon known to occur in the HK lines (Middelkoop et al. 1981). In Table~\ref{uncer}, the observations of HR\,6094 refer to a constant monitoring for 8 consecutive nights. Regarding the youth of the star (member of the UMa moving group), this period is likely to cover a considerable part or the totality of its rotation: indeed, Mayor et al. (2004) estimate its rotational period as 4.7 days, based on the Ca{\small II} H \& K average activity level. So, if the star is not inclined polewards, rotational modulation ought to be seen if H$\alpha$ shows the required sensitivity. As indicated by the small dispersion ($\sigma$ = 0.30), one of the smallest in the table, this is not the case. Rotational modulation, if present, could not be measured.

From this evidence, we see that activity cycles and rotational modulation, both measurable in the HK lines, do not show up in our H$\alpha$ data. The errors committed in the whole calibration - such as the measuring of the width of the chromospheric peak and the computing of the temperatures -, coupled with the lesser sensitivity of H$\alpha$, turn these changes on the net chromospheric fluxes into only minor variations.

If, on the one hand, it obviously represents a loss in accuracy, on the other hand it can be seen as a better characterization of the {\em mean} chromospheric flux, independent of the rotational phase or the activity cycle. This property would be highly desirable, for it would give more reliability when no information about the activity cycle is available. However, it is seen from Fig~\ref{calcium} that the relation between the H$\alpha$ flux and the mean HK flux has considerable scatter.

\begin{table}
\caption[]{Estimate of the uncertainty on the chromospheric fluxes, derived by the fluctuation in several observations of each star. The weighted mean of the distribution of uncertainties is evaluated as $0.45$. {For HR\,6094, the observations refer to a 8-night monitoring in search of evidence of rotational modulation (not found). Fluxes and uncertainties are in units of 10$^{\rm 5}$\, erg\,cm$^{\rm -2}$\,s$^{\rm -1}$.}} 
\label{uncer}
\begin{center}
\begin{tabular}{c c c c}\hline

HR          & No.    & $<F'_{H\alpha}>$ & $\sigma$  \\ \hline
Sun         & 11     & 3.44         & 0.34      \\
6094            & 8      & 7.12         & 0.30      \\
6060            & 6      & 2.93         & 0.52      \\
7373            & 6  & 4.79         & 0.54      \\
5459            & 3      & 4.54         & 0.05      \\
5699            & 3      & 3.21         & 0.25      \\
509         & 3      & 4.74         & 0.66      \\
3538            & 3      & 3.41         & 0.74      \\
1010            & 3  & 3.67         & 0.70      \\
7665            & 3  & 5.50         & 0.86      \\ \hline

\end{tabular}
\end{center}
\end{table}

If the fluctuation in repeated observations is not physical, but statistical, the standard deviations around the means can give estimates of the errors. Assuming this hypothesis, the computed standard deviations are 0.34 for the Sun and 0.52 for HR\,6060. Table~\ref{uncer} shows data for stars for which we have observations separated by more than six months, except for HR\,609 for which the observations are night-to-night. The mean of the standard deviations, weighted by the number of observations, is taken as the empirical uncertainty of the chromospheric fluxes; its value is 0.45, in units of 10$^{\rm 5}$\, erg\,cm$^{\rm -2}$\,s$^{\rm -1}$.

\section{Calibration of the age indicator}
\subsection{Clusters and kinematic groups}

\begin{table}
\caption[]{The stars used to calibrate the age indicator. Effective temperatures in Kelvin; fluxes in 10$^{\rm 5}$\, erg\,cm$^{\rm -2}$\,s$^{\rm -1}$.}
\label{clusters}
\begin{center}

\begin{tabular}{c c c c}
\multicolumn{4}{l}{Pleiades}            \\
\multicolumn{4}{l}{$Age=100\,Myr$}      \\ \hline

HD  & HR    & $T_{\rm{eff}}$& $F'_{H\alpha}$    \\ \hline
282962  & -     & 5903          & 16.85     \\
282975  & - & 5657      & 15.17     \\
BD+23 527&- & 5932      & 13.07     \\ \hline
        &       &               &\\
\end{tabular}

\begin{tabular}{c c c c}
\multicolumn{4}{l}{Ursae Majoris}           \\
\multicolumn{4}{l}{$Age=300\,Myr$}      \\ \hline

HD  & HR    & $T_{\rm{eff}}$& $F'_{H\alpha}$    \\ \hline
11131   & 531b  & 5873      & 12.44     \\
26913   & 1321  & 5990      & 10.46     \\
26923   & 1322  & 5938      & 6.72      \\
39587   & 2047  & 5966      & 10.25     \\
41593   & --    & 5395      & 9.93      \\
147513  & 6094  & 5840      & 6.95      \\
165185  & 6748  & 5876      & 9.44      \\ \hline
        &       &               &\\
\end{tabular}

\begin{tabular}{c c c c}
\multicolumn{4}{l}{Hyades}              \\
\multicolumn{4}{l}{$Age=625\,Myr$}      \\ \hline

HD       &HR    & $T_{\rm{eff}}$& $F'_{H\alpha}$    \\  \hline
1835     & 88   & 5846      & 7.73          \\
27685    & --   & 5759      & 6.80      \\
27859    &--    & 5872      & 9.07      \\
28099    &--    & 5812      & 7.23      \\
28344    &--    & 5835      & 6.91      \\
28992    &  --    & 5903    & 5.57          \\ \hline
        &       &               &\\
\end{tabular}

\begin{tabular}{c c c c}
\multicolumn{4}{l}{HR 1614}             \\
\multicolumn{4}{l}{$Age=2.0\,Gyr$}      \\ \hline

HD  & HR    & $T_{\rm{eff}}$& $F'_{H\alpha}$    \\ \hline
154931  & --    & 5829      & 2.15      \\
161612  & --    & 5587      & 5.19      \\
194640  & --    & 5543      & 4.14      \\ \hline
        &       &               &\\
\end{tabular}

\begin{tabular}{c c c c}
\multicolumn{4}{l}{$\zeta$ Reticuli}        \\
\multicolumn{4}{l}{$Age=3.3\,Gyr$}      \\ \hline

HD  & HR    & $T_{\rm{eff}}$& $F'_{H\alpha}$    \\ \hline
2151    & 98    & 5863      & 0.14      \\
20766   & 1006  & 5701      & 5.68      \\
20807   & 1010  & 5860      & 3.56      \\
196378  & 7875  & 6030      & 2.45      \\ \hline

\end{tabular}

\end{center}
\end{table}

Having constructed the chromospheric diagnostic, we are in a position to calibrate an age indicator using open clusters and SKGs. The stars used belong to the Pleiades and Hyades open clusters, Ursae Majoris, HR\,1614 and $\zeta$ Reticuli SKGs. In Table~\ref{clusters} we list the observed stars, and the ages attributed in the literature. For the Pleiades cluster, the age is 100 Myr --- although there is some controversy in the literature --- and the cluster membership as studied by Schilbach et al. (1995); for the Ursae Majoris group, membership and age are 300 Myr, from Soderblom \& Mayor (1993); for the Hyades, membership and age are 625 Myr, from Perryman et al. (1998); for HR\,1614, the age is 2 Gyr (\cite{feltzing3}), and the membership is as studied by Smith (1983), given that we only became aware of the work by Feltzing \& Holmberg after we had already performed the observations. For $\zeta$ Reticuli, membership was assumed as studied by del Peloso et al. (2000). The age of this group has been revised by us by explicitly taking into account small metallicity differences within the group, and plotting the member stars in the theoretical HR diagrams of Schaerer et al. (1993, and references therein): an average age of 3.3 Gyr is derived.

Of the three stars that compose our sample of the Pleiades, there is evidence that two of them are possibly binaries (\cite{dave3} and private communication). The data are not conclusive, though. In a cluster as distant as the Pleiades, one cannot conclude more than the possibility of a companion. From our data, we notice that HD\,282975 does not show a flux higher than expected, whereas HD\,282962 shows a chromospheric flux almost twice that shown by BD\,+23\,527, a Pleiad with no evidence of binarity. However, it is not clear if it is due to the flux of a companion or just random scatter around the average chromospheric flux of a cluster. Without conclusive evidence of binarity, we did not consider them so.

\begin{figure}
\resizebox{10cm}{7cm}{\includegraphics{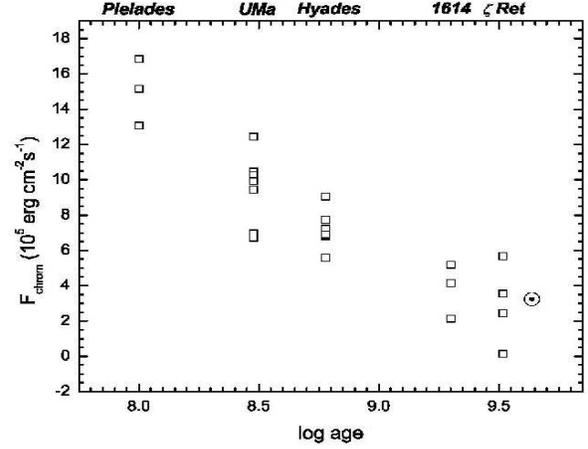}}
\caption[]{Absolute chromospheric flux of the cluster stars and the Sun, plotted against their ages. Even though some scattering is present, the age-activity relation is visible. Skumanich's power law for the Ca\,{\small {II}} K line, of exponent -0.5 (\cite{skumanich}) is substituted here by another one, of exponent (-0.39 $\pm$ 0.01) for H$\alpha$.} 
\label{fluxage} 
\end{figure}

\begin{figure}
\resizebox{10cm}{7cm}{\includegraphics{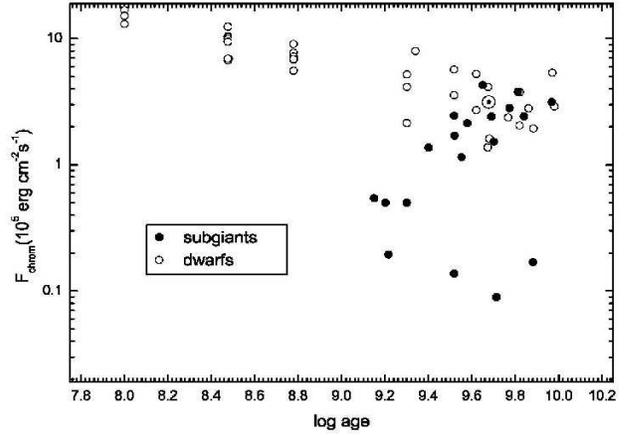}} \caption[]{The stars of our sample for which the solution of age is more accurate than 0.1 dex. The subgiants show a bimodal behaviour, with some acting as low-activity dwarfs, and others not obeying the relation defined by the dwarf stars. These, not following the age-activity relation, are not likely to be magnetically heated.}
\label{subgiants}
\end{figure}

Plotting these data in fig.~\ref{fluxage}, and gathering the cluster points in means, we arrive at the dependence of the chromospheric flux upon age:

\begin{equation}
\label{calib}
\log\, F'_{H\alpha} = 5.79 - 0.39\, \log (age),
\end{equation} where the age is in Gyr. The best fit is calculated considering the error bars, which is needed regarding the great scattering present in the Ursae Majoris and $\zeta$ Reticuli SKGs. Errors are calculated by adding the internal (scattering) and external (0.45) errors in quadrature. The rms of the best fit is 0.01.

\subsection{Field Stars and the Behavior of Subgiants}

Calibration~\ref{calib} is fundamentally defined by young stars, and only the Sun constrains the relation after 4 Gyr. Although there is no reason to believe that the decaying law changes abruptly as the star ages on the main sequence, we want to refine the calibration by inserting stars that sample the age range beyond the Sun. To do so, we rely on field stars, given that stellar clusters are, for the most part, already  spread throughout the Galaxy after such lengths of time. As isochronal ages of field stars are much less accurate, we must select, from our sample, those field stars with parameters determined with maximum accuracy. Although the HIPPARCOS mission reduced the errors in parallax by an order of magnitude, making the errors in parallax almost negligible, uncertainties in temperature are still considerable. Just after the turnoff point isochronal ages are accurate, once the curves assume a horizontal behavior, so errors in temperature do not affect the determination very much.

In this work we determined all temperatures using the wings of H$\alpha$, with a mean uncertainty of 50\,K. For a large fraction of our sample, the literature provided temperatures derived from excitation equilibrium of Fe\,I and Fe\,II lines, and also derived by photometry, using the $(B-V)$, $(b-y)$, $(V-K)$, $\beta$ and the Tycho (B$_{\rm T}$-V$_{\rm T}$) color indices. These temperatures were combined, giving a more accurate value to be used in the isochrones. The isochrones used are those calculated by Schaerer et al. (1993, and references therein), adding the solar zero-point and interpolating to the correct metallicity. By visually inspecting the plots, we discarded stars which were in the ZAMS, in regions were the isochrones overlap, those who gave ages older than presumed to the Galaxy's thick disk ($\sim$ 10 Gyr), or simply those who were in regions where the curves get dense enough to cause uncertainties in age greater than 0.1 dex. Doing so, a large fraction of the stars with a good solution of age, not surprisingly, are subgiants.

Fig~\ref{subgiants} shows the plot for all stars with a solution of age with error smaller than 0.1 dex, separated in dwarfs and subgiants, again using $\log g$ = 4.20 as a boundary. Clearly, dwarfs and subgiants follow distinct populations. A closer look at the figure suggests a bimodal behaviour of the subgiants, with some subgiants behaving as old dwarf field stars, and other ones being much less active. It is concluded that subgiants are powered by a different mechanism, but some of them are still magnetically heated, at least in part, by the same mechanism that heats low-activity dwarfs.

Recent models (Fawzy et al. 2002c, and references therein) claim that, in evolved stars, the efficiency of the dynamo is so drastically reduced that the introduction of magnetic energy in the chromosphere is no longer enough to provoke considerable emission in the opaque frequencies. By these models, the dissipation of acoustic waves should be the main power source, and should be observed as a basal flux, independent of age. If this transition from magnetohydrodynamical to an acoustic mode in fact occurs, we should find that the chromospheric emission from subgiants is not related to age at all. Indeed, the linear regression of the two quantities reveals a null angular coefficient, apparently in support of the model.

\begin{figure}
\resizebox{10cm}{7cm}{\includegraphics{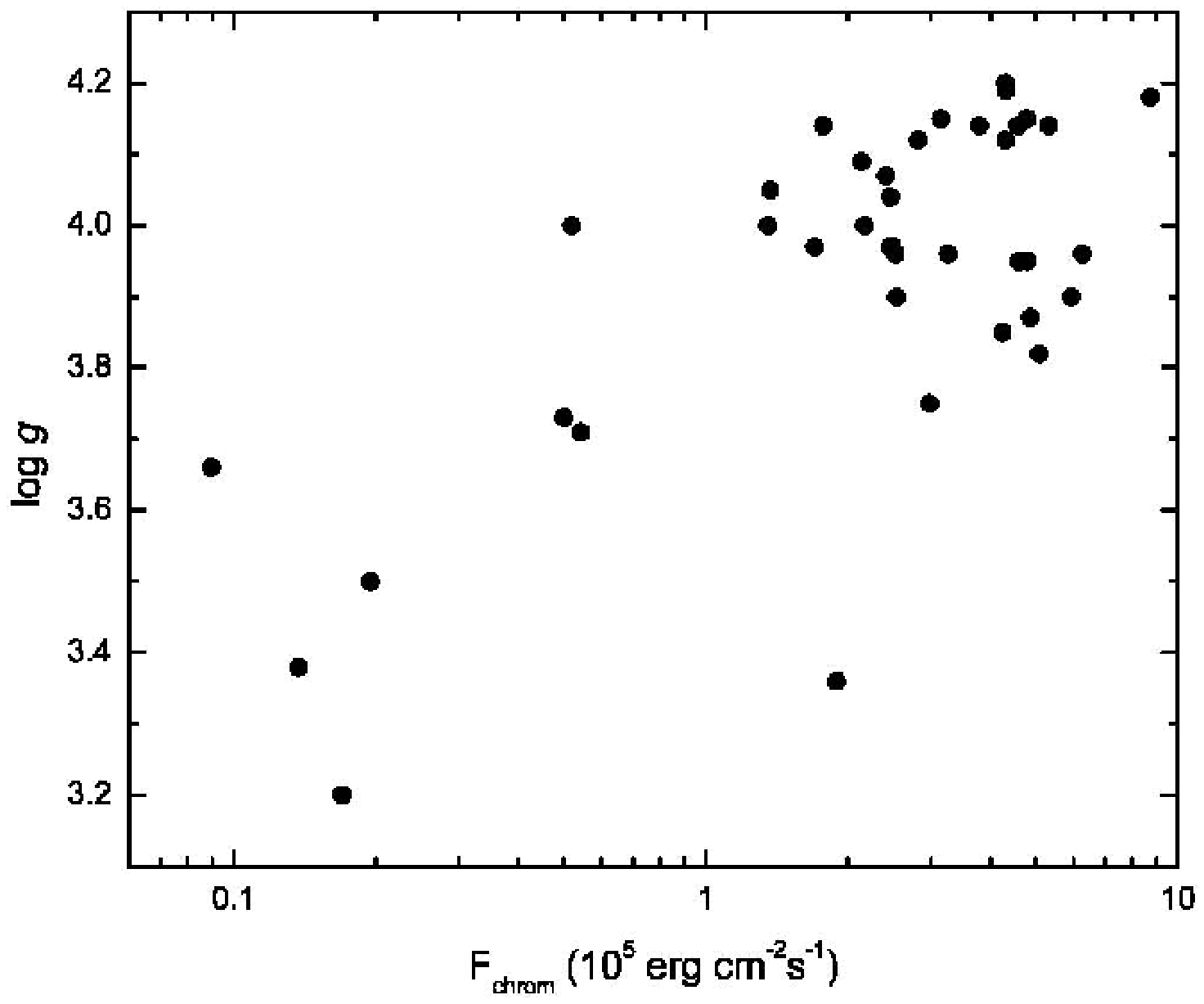}}
\resizebox{10cm}{7cm}{\includegraphics{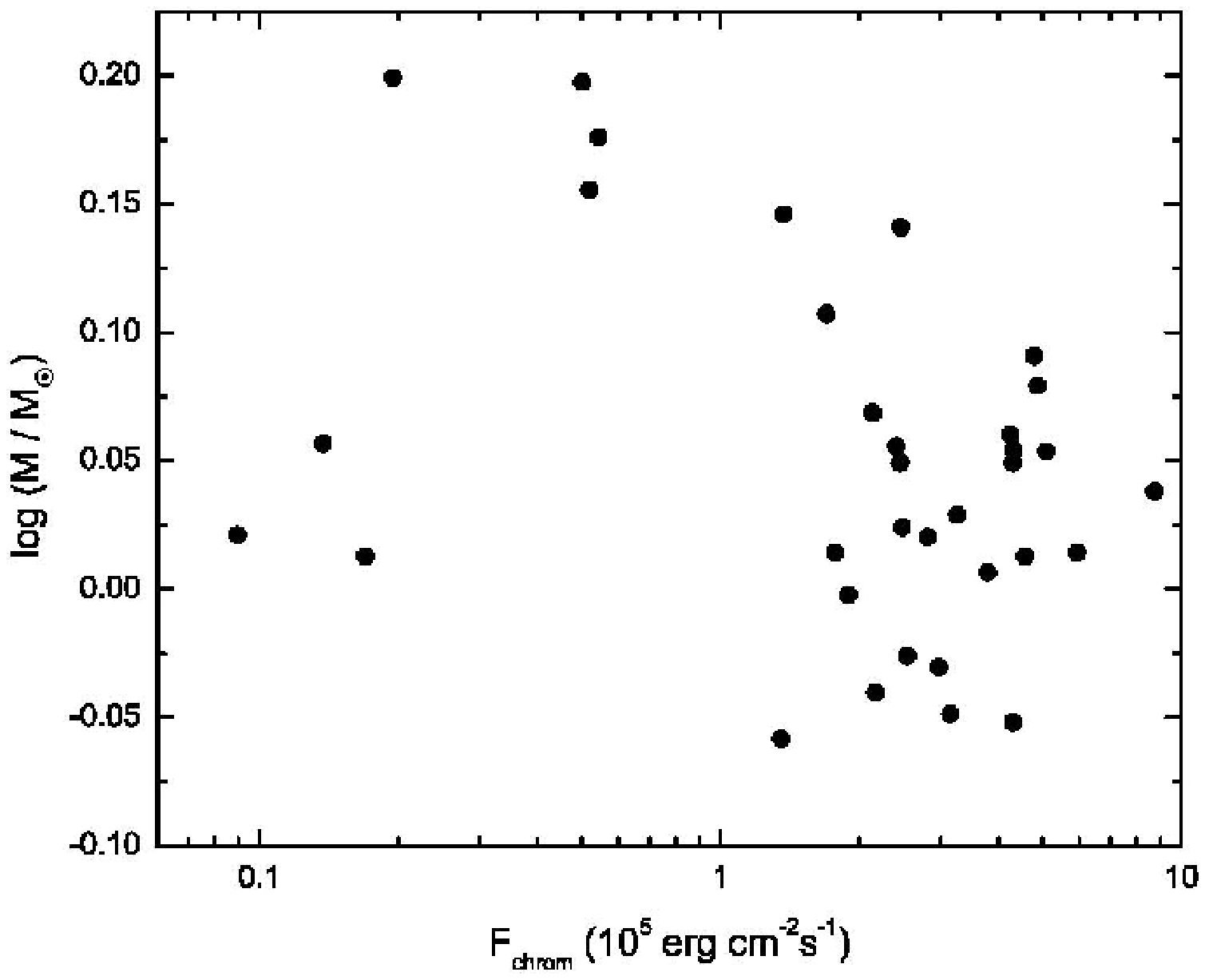}}
\caption[]{Dependence of the chromospheric activity in the subgiant branch upon {\em a}. surface gravity and {\em b}. mass. Only very weak correlations appear, with considerable scatter. The value $\log g$ $\sim$ 3.8 marks the boundary between active and inactive subgiants.}
\label{gravmass}
\end{figure}

Performing a multi-linear regression of the chromospheric flux against age, mass, metallicity, surface gravity, effective temperature and luminosity on the subgiants, we did not find any clear correlation within a $2\sigma$ criteria, but a slight dependence is suggested upon mass and surface gravity, as shown in fig.~\ref{gravmass}.

From the dependence upon gravity, the reason behind the bimodal behavior of the subgiants in fig~\ref{subgiants} becomes clear. The least active stars are those of lowest surface gravity, with the value of $\log g$ $\sim$ 3.8 marking the boundary between active and inactive subgiants. It seems to imply that with decreasing surface gravity (increasing radius), the magnetic phenomena lose importance, the situation moving towards the onset of acoustic heating at $\log g$ $\sim$ 3.8.

Obeying different physics, the subgiants must be discarded from the age indicator. It does not affect the calibration at all, for just two cluster stars were classified as subgiants, following the $\log g <$ 4.20 criteria.

\begin{figure}
\resizebox{10cm}{7cm}{\includegraphics{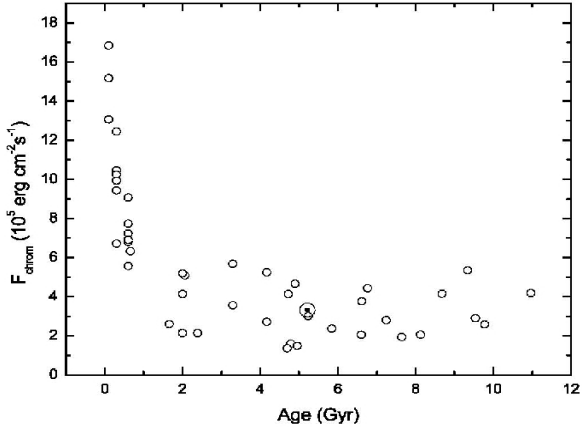}} \caption[]{The age-activity relation for the dwarf ($\log g <$ 4.20) stars of our sample. The curve quickly flattens out after $\sim$ 2 Gyr. }
\label{linageactivity}
\end{figure}

Fig.~\ref{linageactivity} shows a plot, in linear scale, without the subgiants.It is clear that after 2 Gyr the decaying curve flattens and the calibration loses sensitivity: the ages indicated by the degree of chromospheric activity are accurate only for {\em young} stars, losing sensitivity as the star evolves away from the ZAMS. The behavior is then inverse from that of the isochrones, which better discriminate {\em evolved} stars. One thus concludes that the two methods are complementary.

\section{Spicing the dynamo: influence of other parameters in the chromospheric decay}

The spread present in fig.~\ref{linageactivity} and within the clusters and SKGs led us to ask whether there are other parameters influencing the chromospheric activity. Given the physics of the problem, the dynamo effect is the result of a coupling between stellar rotation and convection. Age is only a measurement of the braking the stellar rotation has suffered. Looking at the other ingredient of the dynamo -- convection -- we are naturally led to consider that {\em mass} and {\em metallicity} should play a role in the chromospheric activity: the greater the metallicity, the greater the opacity; plus, the smaller the mass in the main sequence, the closer to the core the radiative energy flux is blanketed. Both  situations have the effect of deepening the convective zone, enhancing the dynamo and, hence, the degree of chromospheric activity. By performing a multilinear regression on the sample of dwarf stars, having chromospheric flux as the dependent parameter, the following results were achieved:

\begin{itemize}
\item Strong anti-correlation with age;
\item Weak anti-correlation with mass;
\item Weak correlation with metallicity;
\item Insensitivity to other parameters.
\end{itemize}

\begin{table}
\label{spice} \caption[]{Multi-linear regression having chromospheric flux as the dependent parameter, and performed only on the dwarf stars of our sample. The parameter $t$ is the ratio of the coefficient by its error. The p-value is the probability-value that the parameter is not significant to explain the chromospheric activity. Age is dominant, but mass and metallicity, rising above $|t| >$ 2, also seem to influence.}
\begin{center}
\begin{tabular}{c c c c c} \hline

Parameter       & Value & Error & t     & p-value   \\ \hline
$\log age$      & -5.55 & 0.77  & -7.21 & $<$0.0001 \\
Mass            & -20.80& 8.74  & -2.38 & 0.02      \\
$[Fe/H]$        & 5.12  & 2.53  & 2.02  & 0.05      \\
$\log g$        & 0.36  & 5.96  & 0.06  & 0.95      \\
$T_{\rm{eff}}$      & 0.006 & 0.004 & 1.5   & 0.15      \\
$\log (L/L_{\sun})$ & 2.53  & 3.67  & 0.69  & 0.50      \\ \hline

\end{tabular}
\end{center}
\end{table}

The full multilinear regression parameters are shown in Table~4. The result is consistent with general predictions of the dynamo model. There is then a strong decline with age along with a second order correction due to metallicity, which, as it increases, makes convection more efficient and enhances chromospheric activity at the same age. Another correction due to mass occurs, which, as it decreases, again makes convection more efficient at the same age.

One might argue, nevertheless, that the dependence upon metallicity could be merely a reflection of the dependence upon age, due to the so called age-metallicity relation in the Galaxy. However, the chemical enrichment is so slow and the galactic disk is so heterogeneous -- in short, the spread in the age-metallicity relation is so severe (Feltzing \& Holmberg 2001) -- that one could easily consider that, for our purposes, the two quantities are independent.

The Sun appears to be an average star, in what pertains to the H$\alpha$ chromospheric flux, for its age, mass and metallicity. This constrasts with the suggestion of Hall \& Lockwood (2000), that HR\,6060, a nearly perfect solar twin (Porto de Mello et al. 1997), has slightly higher Ca\,{\small II} H \& K chromospheric flux than the Sun, as well as a shorter period for its chromospheric cycle and a slightly faster rise from the quiet phase to the active phase, taken overall as a sign of enhanced activity. Adding to the debate, Soderblom (1985) find that the Sun has a typical rotational period for its age; yet Radick et al. (1998) suggest that the Sun has a slightly subdued photometric variability for its degree of chromospheric activity. Probably a more definitive answer to the question of whether the Sun is a typical star for its chromospheric activity will have to await comprehensive analyses of large samples of solar type stars with various spectroscopic activity indicators, calibrated in age, since the different indicators reveal physical information from distinct layers of the chromosphere.

\section{Conclusion}

With high quality data, we developed an accurate chromospheric diagnostic based on the radiative losses in H$\alpha$ using the Barnes-Evans relation (\cite{barnes}) and the photometric waveband $\lambda\lambda$\,6550-6600\,{\AA} defined by Willstrop (1965), using the procedure established by P$^{\rm 2}$91. The precision achieved, with an uncertainty of 0.45 (in units of 10$^{\rm 5}$\, erg\,cm$^{\rm -2}$\,s$^{\rm -1}$) is one of the best in the H$\alpha$ literature. This is due to many reasons, including that the size of the sample used reduced the error inherent to the photospheric subtraction; also, the broader coverage of our spectra, compared to those used by P$^2$91, allowed for a reliable normalization; finally, the sample was observed exclusively with solid state detectors. We also found evidence that the chromospheric emission, as measured in H$\alpha$, is insensitive to the activity cycle and rotational modulation, hence characterizing the mean chromospheric flux in a more adequate way than the Ca\,{\small {II}} K line.

As a byproduct of the analysis, we developed an elegant automated method to determine effective temperatures using the wings of H$\alpha$, structured on the classical K-S test. Tests show that the two methods, visual and automatic, agree to within one standard deviation, which supports its efficiency. The automation of the procedure means, mainly, the suppression of the personal error in the visual procedure, as well as a great economy of time when analyzing several spectra.

With this diagnostic, we calibrated the age-activity relation using stars belonging to open clusters, as well as to SKGs, finding a well defined relation until $\sim$2 Gyr. By inserting field stars with good age solutions, it is found that a significant numer of subgiants do not follow the relation, hence one should not expect their emission mechanism to be magnetohydrodynamical in nature. This result agrees with recent models that credit them as powered by acoustic wave heating, generating basal flux in the low chromosphere.

Regarding the dwarfs, we find that mass and metallicity differences seem to be needed to explain their emission, which is consistent with the dynamo model. It is also verified that the age-activity relation loses sensitivity after 2\,Gyr, flattening beyond this value. The method is then complementary to the isochrones.

For future work, we intend to refine the calibration, inserting more clusters and kinematical groups, and to investigate the spread within the clusters as an effect of mass. Our present data do not allow for this analysis because the individual samples, with a maximum of 6 stars (Hyades and Ursae Majoris) are not statistically significant. In the immediate future, we shall compare the results presented here, with the ones of another work (in preparation), using the Ca\,{\small {II}} H \& K lines. Such a study should contribute both to refining our knowledge of the detailed structure of the chromospheric activity vs. age relationship, and its stratification with stellar parameters, and to provide additional observational constraints to the study of upper stellar atmospheres.

\begin{acknowledgements}
  We thank the referee, Dr.\,R.\,Pallavicini, for very helpful and constructive comments, which considerably improved this paper. This paper is based on the senior thesis of one of the authors (W. Lyra). W. Lyra wishes to thank CNPq for the award of a scholarship, without which this work would not have been done. G. F. Porto de Mello acknowledges financial support bt FAPESP, under grant Projeto Tem\'aticos 00/06769-4 (University of S\~ao Paulo), by CNPq/Brazil under grant 552331/01-5, and by MEGALIT/Instituto do Mil\^enio program. We are extremely grateful to all the staff at Observat\'orio do Pico dos Dias, Brazil, for their unfailing helpfulness and efficiency in the many observational runs needed for this work. This research has made use of the SIMBAD database, operated at CDS, Strasbourg, France.
\end{acknowledgements}

\end{document}